\documentclass[12pt,twoside,fleqn]{article}
\usepackage{epsfig}
\usepackage{psfig}
\def\lsim{\raise0.3ex\hbox{$<$\kern-0.75em\raise-1.1ex\hbox{$\sim$}}}
\def\gsim{\raise0.3ex\hbox{$>$\kern-0.75em\raise-1.1ex\hbox{$\sim$}}}
\def\asymxy{\raise0.3ex\hbox{$\longrightarrow$\kern-3.25em\raise-2.5ex
\hbox{$|\vec{x}-\vec{y}|\rightarrow\infty$}}}
\makeatletter
\@addtoreset{equation}{section}
\makeatother

\newcommand{\beqn} {\begin{equation}}
\newcommand{\eqn} {\end{equation}}
\newcommand{\bc} {\begin{center}}
\newcommand{\ec} {\end{center}}

\newcommand{\slsh}[1] {#1\kern-.43em/}
\newcommand{\real}{{\sf I}\kern-.12em{\sf R}}
\newcommand{\comp}{{\sf I}\kern-.48em{\sf C}}
\newcommand{\nin} {\in\kern-.6em/}

\def\ie{{\sl i.e.\/}}

\def\MEF{m_{\rm eff}}\def\mef{\ifmmode\MEF\else$\MEF$\fi}

\def\SM{s_{\mu}}\def\xm{\ifmmode\SM\else$\SM$\fi}

\begin{document}
\thispagestyle{empty}
%
 \mbox{} \hfill BI-TP 99/04\\
 \mbox{} \hfill {hep-lat/9903030} \\
 \mbox{} \hfill March 1999\\
 \mbox{} \hfill revised version\\
\begin{center}
{{\Large \bf The quenched limit of lattice QCD \\
\vspace*{0.3cm}
at non-zero baryon number}
\bigskip
} \\
\vspace*{1.0cm}
{\large J. Engels, O. Kaczmarek, F. Karsch and E. Laermann} \\
\vspace*{1.0cm}
{\normalsize
$\mbox{}$ {Fakult\"at f\"ur Physik, Universit\"at Bielefeld,
D-33615 Bielefeld, Germany}
}
\end{center}
\setlength{\baselineskip}{1.3\baselineskip}
\vspace*{1.0cm}
\centerline{\large ABSTRACT}

We discuss the thermodynamics of gluons in the background of static
quark sources. In order to do so we formulate the quenched limit of
QCD at non-zero baryon number. A first numerical analysis of this system
shows that it undergoes a smooth deconfining transition. 
We find evidence for a region of coexisting phases that becomes broader 
with increasing baryon number density. Although the action is in our 
formulation explicitly $Z(3)$ symmetric the Polyakov loop expectation value 
becomes non-zero already in the low temperature phase.
It indicates that the heavy quark potential stays finite at large 
distances, \ie~the string between static quarks breaks at non-zero baryon 
number density already in the hadronic phase.

\vspace*{2.0cm}
\noindent
PACS-Indices: 5.70.Ce~~12.38.Gc~~11.30.F 

\vspace*{0.3cm}

\vspace*{-0.2cm}
\hspace*{1.3cm} Deconfinement transition, Quenched limit

\baselineskip 20pt

\noindent

\vskip 20pt
\vfill
\eject
\baselineskip 15pt
\section{Introduction}
 
Numerical studies of the $SU(3)$ gauge theory, \ie~ the heavy quark
mass ({\it quenched}) limit of QCD, are extremely helpful for the 
understanding of the phase structure of QCD at non-zero temperature.
The deconfining phase transition as well as
basic properties of the low and high temperature phases can be
studied in this approximation, although the phase transition in the
light quark mass ({\it chiral}) limit differs not only on a quantitative
level but also qualitatively; in the chiral limit the order of the transition 
is controlled by the chiral symmetry of the fermionic part
of the QCD Lagrangian whereas in the quenched limit the $Z(3)$ center
symmetry dictates the order of the transition. Nonetheless, the fundamental
features of deconfinement which are reflected, for instance, in the release
of many new degrees of freedom at $T_c$, are similar in both limits. This leads
to a quite similar temperature dependence of bulk thermodynamic 
observables like the pressure and energy density \cite{Joswig}. 

So far the investigations of QCD thermodynamics were essentially limited to 
the case of 
vanishing baryon number. In the case of non-zero baryon number, usually
realized in thermodynamic calculations through the introduction of a 
non-zero chemical potential ($\mu$) \cite{Has83,Kog83}, little is known 
from lattice calculations 
about the behaviour of thermodynamic observables, the QCD phase diagram and 
the properties of the different phases.
The reason for this is well-known.
The probabilistic interpretation of the path integral representation
of the QCD partition function breaks down as soon as one introduces
a non-zero chemical potential. Moreover, even the naive quenched limit
at non-zero chemical potential, \ie~the ordinary $SU(3)$
gauge theory at $\mu =0$, turned out to be pathological when fermionic
observables with $\mu \ne 0$ are analyzed \cite{Barxx}. 
In fact, it is understood
that this naive limit is not the correct quenched limit of finite
density QCD; it is the zero flavour limit of a theory with
equal number of fermion flavours carrying baryon number $B$ and $-B$,
respectively. \cite{Stexx}. 

As the relativistic chemical potential $\mu$ also
contains a contribution proportional to the rest mass of the quarks,
the static limit of QCD at non-zero chemical potential requires to take
a double limit in which the quark mass as well as the chemical potential
is taken to infinity while an appropriate ratio is kept fixed. 
This limit has been formulated in \cite{Bender,Blum}. It seems 
that in this case the first order deconfining phase transition of the $SU(3)$ 
gauge theory turns into a crossover for arbitrarily small, non-zero values of 
the chemical potential \cite{Blum}. However, 
although at non-zero chemical potential the thermal phase transition may get 
lost as a true singularity of the partition function of quenched QCD\footnote{This
may even be the case under more general circumstances. If the phase
transition at finite density is more like a percolation transition thermal
observables may be non-singular although singularities due to percolation of
domains with high energy or particle density will still exist 
(Kert\'ez line \cite{Ker89}) \cite{Satz}.}, we still
expect the quenched theory to resemble the physics of deconfinement -- 
gluon thermodynamics in the background of a non-zero number
of static quark sources is expected to lead to deconfinement at high
temperature. It is then interesting to analyze the nature of this transition
and study in how far static quark sources in a gluonic heat bath influence the 
deconfinement of gluons and, for instance, the heavy quark potential. 
We will discuss here a framework for the analysis of these questions -- quenched 
QCD at fixed baryon number -- and will present first numerical simulations
which address these questions. 

Rather than introducing a non-vanishing chemical potential, \ie~formulate
QCD at non-vanishing baryon number density in the grand canonical ensemble,
one may go over to a canonical formulation of the thermodynamics and
fix directly the baryon number \cite{Red87}. 
This is achieved by introducing an imaginary chemical potential \cite{Red87,Weixx}
in the grand canonical partition function. Performing subsequently a 
Fourier integration allows to project onto the canonical 
partition function for a given sector of fixed baryon number \cite{Red87}.
In the heavy quark mass
limit one is then left with a well-defined quenched theory at fixed baryon
number -- gluon thermodynamics in the background of a non-zero number
of static quark sources, suitably arranged to obey Fermi statistics.

Despite the fact that the grand canonical formulation of finite density QCD
leads to severe numerical problems \cite{Bar98} the alternative canonical
approach at non-zero baryon number so far did find only little
attention\footnote{It seems that in addition to the mean-field analysis
performed in Ref.~\cite{Red87} in the heavy quark mass limit so far the
canonical formulation only has been used in a numerical study at strong
coupling ($\beta = 6/g^2 = 0$) using the Monomer-Dimer-Polymer
representation of QCD \cite{MDP}. The potential power of this approach has,
however, also been stressed recently in Ref.~\cite{Alf98}.}.
To some extent this may be justified. Introducing a 
constraint on the total number of fermions (quarks minus anti-quarks)
in general leads to rather complicated non-local constraints for the
gauge field sector. However, in view of the problems that arise in the 
standard non-zero chemical potential formulation and that grow 
exponentially with the size of the
volume one may seriously want to analyze also the canonical approach in more 
detail.  We will show here that the formulation
of QCD at non-zero baryon number does have a well defined, non-trivial
heavy quark mass (quenched) limit which can be analyzed numerically at least for
moderate baryon number densities.  

In the grand canonical formulation the Fermion determinant becomes complex 
for any non-zero chemical potential leading to all the conceptual 
and algorithmic problems. In the canonical approach, on the other hand, the
determinant stays real. However, as should be obvious, the 
problems related to 
a non-positive integration measure are not solved that easily.
The Fourier transformation which is needed in the canonical formulation
to project onto the sector with fixed baryon number reintroduces negative 
contributions to the partition function and one thus again faces algorithmic 
problems. Performing the Fourier integration numerically thus seems to be
ruled out; one can, however, do it explicitly. This leads to a 
complicated expression in terms of products
of quark propagators which may be viewed as the boundary conditions for
the gauge fields needed to project onto a sector with fixed baryon number.
We will use this as a starting point for our analysis of the 
heavy quark mass, quenched, limit.

In the next section we will discuss the canonical formulation of QCD
at finite baryon number and, in particular, introduce the quenched limit.
In section 3 we will present some results of a first numerical analysis
of this quenched theory. Section 4 contains our conclusions. In an appendix
we give explicit representations for the canonical partition function
for some values of the baryon number.

\section{Lattice~ formulation~ of~ QCD~ with~ non-zero baryon number}

The general framework for going from a grand canonical formulation
of QCD at non-zero baryon number density in terms of a non-vanishing
chemical potential \cite{Has83,Kog83} to a canonical formulation with a 
fixed baryon number has been outlined in \cite{Red87}. 
As mentioned before the main difficulty of this approach arises from the
need to perform a Fourier transformation to eliminate the chemical 
potential in favour of a fixed baryon number. We will show in the 
following that this integration can be performed explicitly. The projection
onto a given sector of fixed baryon number is then contained in a 
rather complicated sum over products of quark propagators. This 
representation probably is too complicated to be useful for numerical
calculations with arbitrary, e.g. the physically interesting, 
light quark masses.
However, it may provide a new starting point for physically relevant
approximations to the complete problem. In particular, we will discuss
in the next section the quenched limit at fixed baryon number, \ie~ gluon
thermodynamics in the background of static quarks.

Starting from the standard formulation of QCD at non-zero chemical potential
\cite{Has83} and the corresponding formulation at non-zero baryon number
\cite{Red87} we will rewrite the fermion sector of the lattice action in a 
somewhat more transparent form which makes clear that the non-vanishing
chemical potential can be viewed as a modification of the temporal boundary 
conditions for the fermion fields. 
To be specific we will use the Wilson formulation of the fermion action.
Although we will eventually restrict our discussion to the static limit, 
which also can be obtained directly from the heavy quark formulation for
Wilson fermions derived in analogy to the approach given in Ref.~\cite{Blum}
for staggered fermions, we will start here by formulating the canonical 
approach for arbitrary quark masses.                
 
The action for Wilson fermions at non-zero chemical potential is given by

\begin{eqnarray}
S_F (\mu a) \hspace*{-0.3cm}&\equiv&\hspace*{-0.3cm} \bar{\psi} Q \psi \nonumber \\
\hspace*{-0.3cm}&=&\hspace*{-0.3cm}\sum_x \biggl( \bar{\psi}_x \psi_x  
-\kappa \sum_{j=1}^3
\bigl[ \bar{\psi}_x (1-\gamma_j ) U_{x,j} \psi_{x+\hat{j}} 
+\bar{\psi}_{x+\hat{j}} (1+\gamma_j ) U_{x,j}^{\dagger} \psi_x\bigr]\nonumber \\
& &-\kappa \bigl[{\rm e}^{\mu a} \bar{\psi}_x (1-\gamma_4 ) U_{x,4}
\psi_{x+\hat{4}} + {\rm e}^{-\mu a}
\bar{\psi}_{x+\hat{4}} (1+\gamma_4 ) U_{x,4}^{\dagger} \psi_x\bigr] \biggr)~~.
\label{action}
\end{eqnarray}       

Here $\kappa$ is the hopping parameter which controls the value of the
quark mass, $x=(\vec{x},x_4)$ denotes the sites on a lattice of 
size $N_\sigma^3 \times N_\tau$. The Grassmann fields obey anti-periodic 
boundary conditions in the time direction (fourth direction), 
\ie~ $\psi_{(\vec{x},N_\tau+1)} =-\psi_{(\vec{x},1)}$ and 
$\bar{\psi}_{(\vec{x},N_\tau+1)}=-\bar{\psi}_{(\vec{x},1)}$. 

We may shift the dependence on the chemical
potential to the last time slice, which connects the hyperplanes with $x_4=N_\tau$ 
and $x_4=1$, by transforming the fermion fields\footnote{Note that
the Jacobian of this transformation equals one.} 
\begin{equation}
\psi_{(\vec{x},x_4)}^{\prime} = {\rm e}^{\mu a x_4} \psi_{(\vec{x},x_4)}
\quad , \quad
\bar{\psi}_{(\vec{x},x_4)}^{\prime} = {\rm e}^{-\mu a x_4}
\bar{\psi}_{(\vec{x},x_4)} \qquad .
\label{transformation}
\end{equation}
This eliminates the dependence on the chemical potential on all time
slices but the last one, which now reads

\begin{equation}
\hspace*{-1.0truecm}S_F^{N_\tau} (\mu /T) = \kappa \sum_{\vec{x}} 
\bigl[{\rm e}^{\mu/T} \bar{\psi}_{(\vec{x},N_\tau)} (1-\gamma_4 ) U_{x,4}
\psi_{(\vec{x},1)} + {\rm e}^{-\mu/T}
\bar{\psi}_{(\vec{x},1)} (1+\gamma_4 ) U_{x,4}^{\dagger} 
\psi_{(\vec{x},N_\tau)}\bigr]~,
\label{new_action}
\end{equation}
with $x\equiv (\vec{x},N_\tau)$. We also have used the definition of the 
temperature $1/T= aN_\tau$ in 
writing $\mu/T = \mu aN_\tau$ and furthermore explicitly took 
care of the anti-periodic boundary conditions for the fermions. 
The remaining part of the fermion action, which now is independent of the
chemical potential, may be written as

\begin{equation}
\tilde{S}_F = S_F (0) - S_F^{N_\tau} (0) \quad .
\label{rest_action}
\end{equation}

This representation explicitly shows that the formulation of 
thermodynamics with non-zero chemical potential can be viewed as a 
generalization of the non-zero temperature case, which is realized through 
anti-periodic boundary conditions in the fermion sector. They are now 
generalized to 
$\psi_{(\vec{x},N_\tau+1)}= -\exp{(\mu/T)}~\psi_{(\vec{x},1)}$
and $\bar{\psi}_{(\vec{x},N_\tau+1)}= -
\exp{(-\mu/T)}~\bar{\psi}_{(\vec{x},1)}$.

So far we have only re-organized the various terms in the fermion sector
of the QCD partition function where we have used the standard Wilson
formulation. For the gluon sector we do not have to specify at this point
the explicit form of the action, $S_G$. 
The partition function in a volume $V= (N_\sigma a)^3$
at temperature $T=1/N_\tau a$ and non-zero chemical potential $\mu a$ then
reads, 
\begin{equation}
Z(\mu/T,T,V) = \int \prod_{x,\nu} {\rm d}U_{x,\nu} \prod_{x} {\rm d}\bar{\psi}_{x}
{\rm d}\psi_x {\rm e}^{-S_F^{N_\tau} (\mu/T)} {\rm e}^{-S_G-\tilde{S}_F}~~.
\label{partition_function}
\end{equation} 
We want to use this form as a starting point to go over to a 
formulation at non-zero baryon number rather than at non-zero chemical 
potential.
This can be achieved by introducing a {\it complex chemical potential} 
and performing a Fourier transformation,
\begin{equation}
Z(B,T,V) = {1\over 2\pi} 
\int_0^{2\pi}  {\rm d}\phi\ {\rm e}^{-iB\phi}\ Z(i\phi,T,V)~~,
\label{density}
\end{equation}
where $B$ denotes the quark number, \ie~the baryon number equals $B/3$.
The Fourier transformation only operates on the factor
${\rm e}^{-S_F^{N_\tau}}$ which only involves links pointing in the 4th 
direction on the last time slice of the lattice. Making use of 
the Grassmann properties of the fermion fields this contribution
can be written as 
\begin{eqnarray}
\hspace*{-0.8truecm}{\rm e}^{-S_F^{N_\tau} (i\phi)} = 
\prod_{(\vec{x},a,b,\alpha,\beta,f)}& & \hspace*{-1.2cm}\biggl(1 - \kappa
{\rm e}^{i\phi} \bar{\psi}_{(\vec{x},N_\tau)}^{a,\alpha,f}  
{\cal U}_{\vec{x}}^{a,\alpha,b,\beta}
\psi_{(\vec{x},1)}^{b,\beta,f}\biggr)\times \nonumber \\ 
\prod_{(\vec{x},a,b,\alpha,\beta,f)}& &\hspace*{-1.2cm}
\biggl(1 - \kappa {\rm e}^{-i\phi}
\bar{\psi}_{(\vec{x},1)}^{a,\alpha,f} {\cal U}_{\vec{x}}^{\dagger a,\alpha,b,\beta} 
\psi_{(\vec{x},N_\tau)}^{b,\beta,f} \biggr)~ ,
\label{contribution}
\end{eqnarray}
where the product runs over all possible combinations of indices with
$\vec{x}$ taking values on the three dimensional (spatial) lattice of 
size $N_\sigma^3$. We also have introduced the notation
${\cal U}_{\vec{x}}=\Gamma_{-} U_{(\vec{x},N_\tau),4}$ and  
${\cal U^{\dagger}}_{\vec{x}}=\Gamma_{+} U_{(\vec{x},N_\tau),4}^{\dagger}$  
with $\Gamma_{\pm} = (1 \pm \gamma_4 )$. Note that the fields
${\cal U}$ carry spinor and color indices which we denote by Greek 
and Latin letters, respectively. We will combine these to a single index
denoted by, e.g. ${\cal A}=(\alpha,a)$ with $\alpha = 1,..., 4$
and $a=1,~2,~3$. In addition we also have allowed for different fermion flavours,
$f=1,..., n_f$ but ignored the possibility of having different quark masses,
\ie~different hopping parameters $\kappa$ for the various flavours.

We may expand the product appearing in Eq.~\ref{contribution} and write it as 
a series in terms of the complex fugacity, $z=\exp (i\phi)$. The Fourier 
transformation in Eq.~\ref{density} will receive a non-zero contribution only 
from the term proportional to $z^B$. The coefficient
of $z^B$ will receive contributions from $n$ terms proportional to $z$
and $\bar{n}$ terms proportional to $z^*$ where $B=n-\bar{n}$. 
Each such contribution is proportional to $\kappa^{n+\bar{n}}$. This 
is the basis for a systematic hopping parameter expansion for the boundary
term. Clearly we need at least $B$ terms proportional to $z$. 
The leading contribution thus arises from the $\bar{n}\equiv 0$ sector.
It can be summarized as 
\begin{equation}
z^B f_B \equiv (-z \kappa)^{B} \sum_{X,{\cal C,D},F}
\prod_{i=1}^{B}
\bar{\psi}_{(\vec{x}_i,N_\tau)}^{{\cal C}_i,f_i}  
{\cal U}_{\vec{x}_i}^{{\cal C}_i,{\cal D}_i}
\psi_{(\vec{x}_i,1)}^{{\cal D}_i,f_i} ~~,
\label{total_B}
\end{equation}
where $X,~{\cal C,~D},~F$ are $B$-dimensional vectors, \ie~ 
$X=(\vec{x}_1,...,\vec{x}_B)$, $F=(f_1,...,f_B)$ and so on.
Of course, all elements of the set of indices 
$\{({\cal C}_i,f_i,\vec{x}_i)\}_{i=1}^B$ as well as 
$\{({\cal D}_i,f_i,\vec{x}_i)\}_{i=1}^B$ have to be different to give a 
non-vanishing contribution to the sum in Eq.~\ref{total_B}. 
The Fourier integral in Eq.~\ref{density} can then be performed explicitly
and we obtain for the partition function at fixed baryon (or quark) number

\begin{equation}
Z(B,T,V) = 
\int \prod_{x,\nu} {\rm d}U_{x,\nu} \prod_{x} {\rm d}\bar{\psi}_{x}
{\rm d}\psi_x f_B~{\rm e}^{-S_G-\tilde{S}_F}~~.
\label{density_B}
\end{equation}
 
To leading order in the hopping parameter the projection onto a sector with 
fixed quark number $B$ is encoded in the function $f_B$ which is a sum
over products of quark propagators between the time slices at $x_4=1$
and $x_4=N_\tau$. In higher orders one will, of course, also get 
contributions from quarks propagating backward in Euclidean time. 
In fact, if we think in terms of a hopping parameter expansion 
(heavy quark mass limit) for the entire fermion determinant, the function
$f_B$ is all we need to generate
the leading contribution, which finally will be ${\cal O} (\kappa^{BN_\tau})$.
Higher order contributions will result from $\phi$-independent terms 
coming from an expansion of
$\exp (-\tilde{S}_F)$ as well as from additional factors in the expansion
of Eq.~\ref{contribution} which then have to contain an equal number of 
additional backward and forward propagating terms.

Let us look in more detail at the leading contribution arising from $f_B$. 
For this it is
convenient to evaluate $f_B$, which of course is a gauge invariant
function, in a specific gauge. Let us perform a gauge transformation such
that all the links pointing in the time direction on the last time slice
are equal to unity. This gives
\begin{equation}
f_B =(-2 \kappa)^{B} \sum_{X,{\cal A},F}
\prod_{i=1}^{B}
\bar{\psi}_{(\vec{x}_i,N_\tau)}^{{\cal A}_i,f_i} 
\psi_{(\vec{x}_i,1)}^{{\cal A}_i,f_i} ~~.
\label{f_B}
\end{equation}
Like in Eq.~\ref{total_B} the vectors $X,~{\cal A},~F$ are of length $B$. 
However, now the spinor indices $\alpha_i$ which are part of ${\cal A}_i$ 
only take on the values $\alpha_i=1,~2$ because only the two
components of $\Gamma_-$ are non-zero. This also gives rise to the
factors of $2$ in front of $\kappa$.
When evaluating the Grassmann integrals each of the $\bar{\psi}$ terms
can be contracted with all those $\psi$ terms which carry the same flavour
index. Each pair gives rise
to a matrix element of the inverse of $\tilde{Q}$, 
the fermion matrix corresponding to 
$\tilde{S}_F$. The different pairings give
rise to the Matthews-Salam determinant \cite{Matt}. 
We thus will get the product of $n_f$
determinants, each of dimension $d_l$ such that $\sum_{l=1}^{n_f} d_l = B$,
\begin{equation}
\hat{f}_B =(2 \kappa)^{B} \sum_{X,{\cal A},F}
\prod_{l=1}^{n_f}
{\rm det} {\cal M}_l[x,{\cal A}]~~,
\label{det_B}
\end{equation} 
where the matrix ${\cal M}_l$ gives the contributions for the l-th flavour
and the matrix elements are the corresponding quark propagators, 
\begin{equation}
{\cal M}_l^{i,j} = \tilde{Q}^{-1}_{((\vec{x}_j,1),{\cal A}_j),
((\vec{x}_i,N_\tau),{\cal A}_i)} ~~.
\label{matrix}
\end{equation}
Each matrix element of ${\cal M}_l$ is 
${\cal O}(\kappa^{(N_\tau-1 +|\vec{x}_i -\vec{x}_j|)})$. In the limit of small
$\kappa$-values, \ie~ in the heavy quark mass limit, only matrix elements 
of ${\cal M}$ 
with $|\vec{x}_i -\vec{x}_j|=0$ will thus survive. In this case the elements of 
$\tilde{Q}^{-1}_{((\vec{x}_i,1),{\cal A}_j),((\vec{x}_i,N_\tau),{\cal A}_i)}$ are 
just products of
terms $\Gamma_- U_{(\vec{x}_i,k),4}$ with $k=1,..,N_\tau-1$. As $\Gamma_-$ is
a diagonal matrix in spinor space the indices $\alpha_i$ and $\alpha_j$ have 
to be identical. The 
spinor part thus gives rise to an overall factor $2^{N_\tau-1}$
for each $i$, \ie~we obtain $B$ such factors. The multiplication of the SU(3)
matrices yields an element of the ordinary, complex valued (!) Polyakov 
loop ($U\equiv 1$ on the last time slice !) which we denote 
by $L_{\vec{x}_i}^{a_i,a_j}$. Finally, the sum over different colour indices
appearing in Eq.~\ref{det_B} leads to contributions 
involving only traces over powers of the Polyakov loop, 
\begin{equation}
L_{\vec{x}} = \prod_{x_4=1}^{N_\tau} U_{(\vec{x},x_4)}~~.
\label{poly}
\end{equation} 

As the (colour, spinor) label ${\cal A}_i$ can take on six different 
values the
determinant is non-zero only if at most six quarks of a given flavour
occupy a given site $\vec{x}_i$. At most three quarks can have the same
spinor component. There are thus six possible contributions, $D_n$, of a given 
site to the determinant depending on the quark occupation number, $n$, of this 
site. These can be expressed in terms of three functions $M_i$ which 
correspond to the number of quarks ($i$) with identical spinor components. 
These contributions are

\begin{eqnarray}
D_1 &=& 2 M_1 \nonumber \\
D_2 &=& 2 (M_1^2+  M_2) \nonumber \\
D_3 &=& 2(3 M_1 M_2+  M_3) \nonumber \\
D_4 &=& 2 (4 M_1 M_3+ 3 M_2^2) \nonumber \\
D_5 &=& 20 M_2 M_3 \nonumber \\
D_6 &=& 20 M_3^2   \quad ,
\label{Di}
\end{eqnarray}
with
\begin{equation}
M_1= {\rm Tr}L_{\vec{x}_i} \quad,\quad
M_2=({\rm Tr}L_{\vec{x}_i})^2 - {\rm Tr}L_{\vec{x}_i}^2\quad, \quad
M_3=6 \quad .
\end{equation}
We note that this 
representation is quite similar to that of the heavy quark mass limit for 
QCD with staggered fermions at non-zero chemical potential \cite{Blum}.

We now may write $\hat{f}_B$ as
\begin{eqnarray}
\hat{f}_B &=& (2 \kappa)^{B N_\tau}\sum_{X,F} 
\prod_{k=1}^{B^{\prime}} 
\prod_{l=1}^{n_f}
D_{n_{k,l}} (\vec{x}_k) \quad ,
\label{f_B_strong}
\end{eqnarray}
where $B^{\prime} \le B$ is the number of distinct sites $\vec{x}_k$ appearing
in $X$ and $n_{k,l}$ is the {\it occupation number} 
for these distinct sites $\vec{x}_k$ with quarks of flavour $l$. They obey the
constraint $\sum_{k,l} n_{k,l} = B$ with $0 \le n_{k,l} \le 6$. The total 
number of quarks at the
site $\vec{x}_k$ is $n_k=\sum_l n_{k,l}$. The sum over all different 
distributions of the flavour number can be performed locally for a given 
distribution of sites $X$. Let us define

\begin{equation}
\bar{D}_{n_k} (\vec{x}_k) =
\sum_{n_{k,1}=0}^{\bar{n}_k} ...\sum_{n_{k,n_f}=0}^{\bar{n}_k}
\delta(n_k- \sum_{l=1}^{n_f} n_{k,l})
\prod_{l=1}^{n_f} D_{n_{k,l}} (\vec{x}_k) \quad ,
\end{equation}
with $\bar{n}_k={\rm min}(n_k,6)$. 
We then can write Eq.~\ref{f_B_strong} as
\begin{eqnarray}
\hat{f}_B &=& (2 \kappa)^{B N_\tau}\sum_{X}
\prod_{k=1}^{B^{\prime}}
\bar{D}_{n_k} (\vec{x}_k) \quad .
\label{f_B_strong2}
\end{eqnarray}

One of the difficulties with using a representation like Eq.~\ref{f_B_strong}
or Eq.~\ref{f_B_strong2} in a numerical calculation
is that there appears a $B$-fold sum over the volume, $V$, with constraints on
the occupation number for a given site, \ie~the computational effort would
be ${\cal O} (V^B)$. Through a cluster decomposition we can, however, reduce 
this to
the evaluation of certain moments of $\bar{D}_i$ as well as their average 
on a given configuration. This is described in the Appendix.

We finally obtain
\begin{eqnarray}
\hspace*{-1.0cm}
\hat{f}_B &=& (2 \kappa)^{B N_\tau} 
\sum_{\{ g_\alpha \} } \delta \biggr( B - \sum_{l=1}^{6n_f} \sum_{\{ \alpha \} } l 
g_\alpha a_l  
\biggl) \times \nonumber \\
& &
\prod_{\{ \alpha \} } {1\over g_\alpha!} 
\biggl\{ (-1)^{(\sum_{l=1}^{6n_f} a_l-1)} 
\biggl(  (\sum_{l=1}^{6n_f} a_l -1)! 
\biggl[ \prod_{l=1}^{6n_f} {1\over a_l! } \bar{D}_l^{a_l}\biggr] 
\biggr)  \biggr\}^{g_\alpha} \quad ,
\label{partitions}
\end{eqnarray}
where $[...]$ is the mean value defined in (\ref{brackets}) and the
product runs over all
possible sets of vectors $\alpha=(a_1,...,a_{6n_f})$ which are constrained by 
\begin{equation}
\sum_{l=1}^{6n_f} l a_l  \le B   \quad .
\end{equation}
The sum over $g_\alpha$ is defined in (\ref{gsumb}).

Now we have achieved that only simple sums over products of suitably
chosen terms $\bar{D}_n$ have to be evaluated to obtain the contribution
$\hat{f}_B$ to the partition function which results
from a given number of static quarks.  
The number of possible partitions contributing to 
Eq.~\ref{partitions} grows, however, rapidly with $B$. 
After all it is nothing else but the result of an 
explicit evaluation of the fermion determinant in a fixed baryon number 
sector. As $\hat{f}_B$ contains products of mean values calculated on a
given configuration, it is a non-local quantity which in a Monte-Carlo
simulation has to be evaluated for each update of a temporal link.
In practice,
calculations thus will be limited to small values of $B$. Some explicit
representations of $\hat{f}_B$ for small values of $B$ are given
in the appendix. 
In the next section we will use this representation
of the quenched limit of the QCD partition function with fixed 
baryon number. We will perform first exploratory Monte-Carlo simulations
to analyze the phase structure of gluon thermodynamics in the presence
of static quarks.

\section{Simulation~ of~ quenched~ QCD~ with~ non-zero baryon number}

For any fixed value of the baryon number we can write the quenched
partition function as   
\begin{equation}
Z(B,T,V) = \int \prod_{x,\nu} {\rm d}U_{x,\nu} ~\hat{f}_B~ {\rm e}^{-S_G}~~,
\label{density_quench}
\end{equation}
where the constraint on the baryon number is encoded in the function
$\hat{f}_B$ given in Eq.~\ref{f_B_strong}. In particular, we note that the 
global $Z(3)$ symmetry of the QCD partition function at non-zero baryon number
is preserved also in the quenched limit, \ie~the 
function $\hat{f}_B$ is invariant under global $Z(3)$ transformations if
$B$ is a multiple of $3$. As the gluonic action $S_G$ also shares this
property the partition function, $Z(B,T,V)$, is non-zero only if $B$ 
is a multiple of $3$.  

We also note that $\hat{f}_B$ is still a complex function. However, upon
integration over the gauge fields the imaginary contribution vanishes;
the partition function is real, of course. Actual calculations thus 
can be performed with ${\rm Re} \hat{f}_B$. The crucial question for a
simulation with this partition function is whether the remaining sign 
changes of the
real part of $\hat{f}_B$ are seldom enough so that a Monte-Carlo 
simulation can be performed with the absolute value of ${\rm Re} \hat{f}_B$
and the overall sign can be included in the calculation of 
averages \cite{Engels}.

\begin{figure}[htbp]
  \begin{center}
    \leavevmode
    \epsfig{file=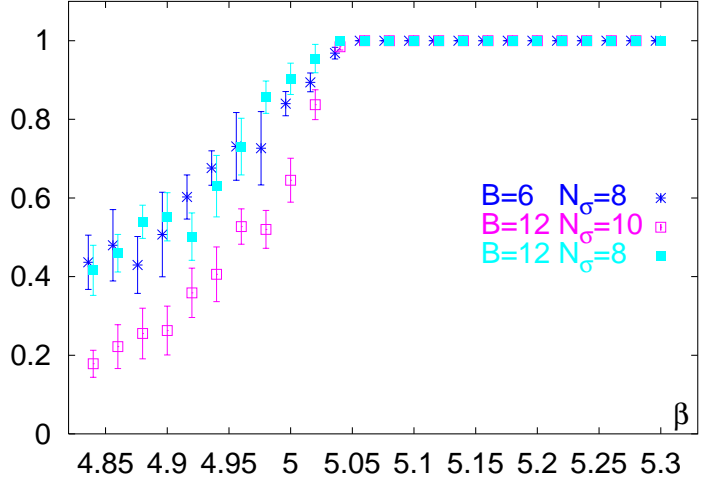, width=100mm}
  \end{center}
  \vspace*{-5ex}
  \caption{Expectation value of the sign of ${\rm Re} \hat{f}_B$,
$\langle {\rm sgn(Re}\hat{f}_B)\rangle_{||}$,
for $B=6$ and 12 and lattices of size $N_\sigma^3 \times 2$ with  
$N_\sigma =8$ and 10.}
  \label{fig:sign}
\end{figure}

We will perform simulations for the one flavour case $(n_f=1)$ using the 
partition function

\begin{equation}
Z_{||}(B,T,V) = \int \prod_{x,\nu} ~{\rm d}U_{x,\nu}|{\rm Re} \hat{f}_B|~ 
{\rm e}^{-S_G}~~.
\label{quench_abs}
\end{equation}
Expectation values of an observable ${\cal O}$ calculated with the
Boltzmann weights defined by the partition function $Z(B,T,V)$ will therefore
be calculated according to

\begin{equation}
\langle {\cal O} \rangle = {\langle {\cal O}
\cdot{\rm sgn} ({\rm Re} \hat{f}_B)  
\rangle_{||} \over \langle {\rm sgn} ({\rm Re} \hat{f}_B) \rangle_{||} }~~.
\label{average}
\end{equation}
Our simulations are performed on lattices of 
size $8^3\times2$ and $10^3\times 2$  using the standard Wilson gauge 
action \cite{Wilson}. For the link updates we use a Metropolis algorithm.
For each link update the change in the function ${\rm Re}
\hat{f}_B$ is calculated and a possible change in sign is monitored.
In all the cases we have studied so far we find that 
$\langle {\rm sgn} ({\rm Re} \hat{f}_B) \rangle_{||}$ is large and can 
be well determined. In fact, for large values of the temperature 
${\rm sgn} ({\rm Re} \hat{f}_B)$ is almost always positive. 
This is evident from Figure~\ref{fig:sign} which shows the average
sign as a function of the coupling $\beta$. The expectation value
of ${\rm sgn} ({\rm Re} \hat{f}_B)$ mainly depends on the spatial
volume $N_\sigma$ but varies little with $B$ at fixed $N_\sigma$.
We also find that
the values of observables like the average action or the Polyakov loop 
do not depend much on the sign of ${\rm Re} \hat{f}_B$ so that the 
errors obtained for these observables from a jackknife analysis
are substantially smaller than those shown in Figure~\ref{fig:sign}. 

A numerical analysis of the thermodynamics at fixed baryon number, $B$, 
can closely follow the standard approach at $B=0$, \ie~in a pure
$SU(3)$ gauge theory \cite{Boyd}. We may analyze the temperature 
dependence of bulk thermodynamics, the Polyakov loop expectation value
and other observables for a gluon gas in the background of static 
quarks. We started a first exploratory analysis of this system by
performing a numerical simulation on lattices with temporal extent $N_\tau=2$.
The simulations have been carried out in the vicinity of the
critical coupling for the deconfinement transition at $B=0$, \ie~for 
gauge couplings $\beta \simeq 5.0$ which are still in the strong coupling 
regime. 
Calculations with fixed $B$ are performed
on lattices of size $N_\sigma^3 \times N_\tau$ and the temperature is 
varied, as usual, by changing the coupling $\beta= 6/g^2$. The dimensionless
parameter kept fixed in the simulation thus is the baryon number density 
in units of the temperature cubed,

\begin{equation}
{n_B \over T^3} = {B\over 3} \biggl( {N_\tau \over N_\sigma} \biggr)^3~~.
\label{fixed_parameter}
\end{equation}
The baryon number density in physical units thus is
\begin{equation}
n_B = {B\over 3} \biggl( {N_\tau \over N_\sigma} \biggr)^3 
\biggl( {T \over 200~{\rm MeV}} \biggr)^3 {\rm fm}^{-3}~~.
\label{physical_density}
\end{equation}
For orientation we note that close to $T_c$, which for the $SU(3)$ gauge
theory is known to be about $270$~MeV, a simulation on an $8^3\times 2$
lattice with $B=12$ corresponds to $n_B \simeq 0.15/{\rm fm}^3$, 
\ie~approximately nuclear matter density.

For vanishing baryon number the Polyakov loop expectation value or 
more precisely the expectation value of its normalized absolute value calculated 
on a finite lattice, 
\begin{equation}
\langle |L| \rangle_V = \langle \bigl|{1\over 3 N_\sigma^3} \sum_{\vec{x}}
{\rm Tr} L_{\vec{x}}\bigr| \rangle~~,
\label{polyexp}
\end{equation}
is an order parameter for the deconfinement transition 
in the infinite volume limit,
\begin{equation}
\langle L \rangle \equiv \lim_{N_\sigma \rightarrow \infty} 
\langle |L| \rangle_V~~.
\end{equation}
As the phase transition is first order 
in $SU(3)$, $\langle L \rangle$ 
changes discontinuously at $T_c$. Besides being related to 
the spontaneous breaking of the global $Z(3)$ symmetry of the pure gauge
action the behaviour of the Polyakov loop also reflects the large distance
behaviour of the heavy quark potential

\begin{equation}
e^{-V(\vec{x}-\vec{y},T)/T}~=~ 
\langle {\rm Tr} L_{\vec{x}} {\rm Tr} L^{\dagger}_{\vec{y}} \rangle \qquad 
\asymxy\quad
9 |\langle L \rangle|^2 \quad .
\end{equation}
The vanishing of $\langle L \rangle$ indicates that the heavy quark 
potential is confining at large distances, while a finite value 
of $\langle L \rangle$ shows that the potential stays finite for infinite
separation of the quark anti-quark pair. In QCD with dynamical light quarks 
the Polyakov loop is no longer an order parameter. The heavy quark
potential stays finite at large distances even in the confined phase
because the static quark anti-quark pair can be screened through the 
creation of a light quark anti-quark pair from the 
vacuum ({\it string breaking}).
For light enough quarks this is indeed observed in numerical simulations at
vanishing baryon number density \cite{DeTar}. 

\begin{figure}[htbp]
  \begin{center}
    \leavevmode
    \epsfig{file=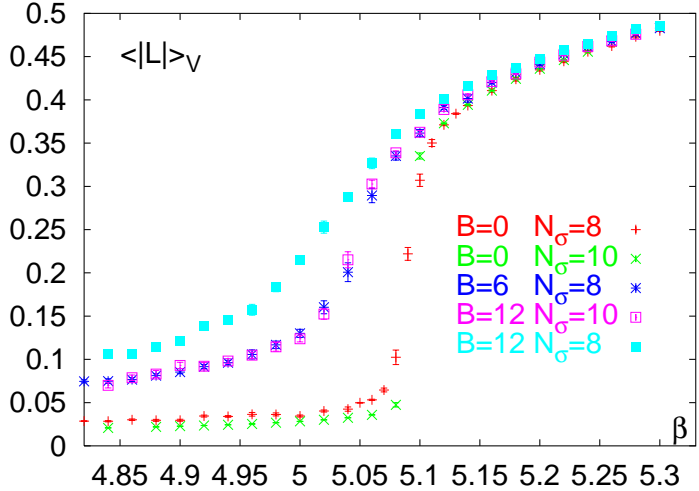, width=100mm}
  \end{center}
  \vspace*{-5ex}
  \caption{Polyakov loop expectation value for different values of $B$
and lattices with spatial extent $N_\sigma =8$ and 10.}
  \label{fig:Polyakov}
\end{figure}

At non-zero baryon number density we expect to find a similar behaviour of
the heavy quark potential even in the heavy quark mass limit because the
quarks needed to break the string need not be created
through thermal (or vacuum) fluctuations. The static quark anti-quark sources 
used to probe the heavy quark potential can recombine with the already present
static quarks and will lead to string breaking even in the low temperature
hadronic phase. We thus expect that the Polyakov loop expectation value 
will not be an order parameter, although the integrand of the partition 
function, $\hat{f}_B \exp (-S_G)$, is $Z(3)$ symmetric, \ie~we expect that  
\begin{equation}
\langle L \rangle  
~~>~~ 0\quad,\quad {\rm for~all}~n_B > 0 ~~{\rm and~all}~~\beta \ge 0~~. 
\end{equation}
This is indeed evident from the results obtained for the Polyakov loop 
expectation values from our simulations with $B=6$ and 12 
on $8^3 \times 2$ and $10^3 \times 2$ lattices which are shown in 
Figure~\ref{fig:Polyakov}. 
We thus find first (indirect) evidence for the modification of the long
distance part of the
heavy quark potential in nuclear matter. This will be analyzed in more 
detail in the future. We also note that there is no significant volume 
dependence at fixed $n_B$\footnote{Calculations performed on a  
$8^3 \times 2$ lattice with $B=6$ and on a $10^3 \times 2$ lattice with $B=12$ 
are performed at nearly the same baryon number density, \ie~
$n_B/T^3 = 0.03125$ and $n_B/T^3 = 0.032$, respectively.}.
This also shows that in 
the thermodynamic limit the physical observables will indeed only depend
on the density rather than the baryon number itself. 

\begin{figure}[htbp]
  \begin{center}
    \leavevmode
\hspace{-0.5cm}
    \epsfig{file=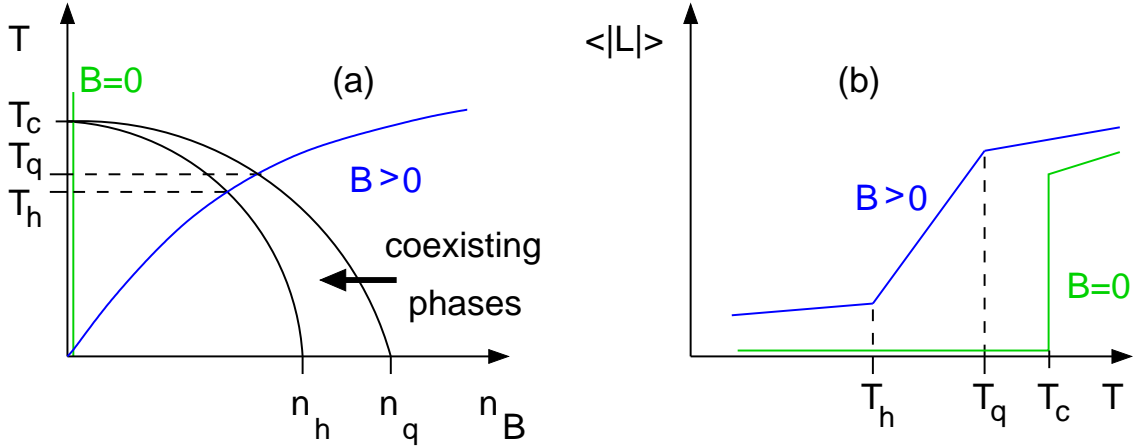, width=150mm}
  \end{center}
  \vspace*{-5ex}
  \caption{Schematic plot of the QCD phase diagram (a) in the 
temperature-baryon number density plane for the case of first order 
transitions in the
entire plane. For $B>0$ and $T_h<T<T_q$ the system stays in a region of  
two coexisting phases.
For $B=0$ the transition occurs at a unique temperature $T_c$.
In (a) we also show the paths followed when varying the
coupling $\beta$ in a Monte-Carlo simulation with fixed $B$, $N_\sigma$
and $N_\tau$. In (b) the expected behaviour of the Polyakov
loop expectation value along these paths of non-zero $B$ as well as for $B=0$
is shown.}
  \label{fig:schematic}
\end{figure}

For $B=0$ there is a clear signal for a first order phase
transition which leads to a discontinuity in $\langle L \rangle$. For all 
$B >0$ we clearly observe a transition from a low temperature phase with small
Polyakov loop expectation value to the high temperature regime characterized 
by a large Polyakov loop expectation value, which is similar to that of the
$B=0$ case. The transition occurs in a temperature interval that broadens
with increasing baryon number density. There is no indication for
a discontinuous transition. In fact, this is not to be expected
in a canonical calculation, even if the transition is first order.
By changing the gauge coupling $\beta$ we vary the lattice cut-off and
through this also the baryon number density continuously. At fixed non-zero 
baryon number
we therefore follow a simulation path that traverses continuously through
a region of two coexisting phases. This situation is schematically 
illustrated in Figure~\ref{fig:schematic}. 

The question now is whether the transition region really is a 
region of coexisting phases.
In this case the values of 
thermodynamic observables result as the superposition of
contributions from two different phases  appropriately weighted 
by the fraction each phase contributes in the coexistence regime.
In the infinite volume limit this could  result in a discontinuous 
change of the slope of thermodynamic observables 
when entering and leaving the coexistence region 
(for an illustration see e.g. Figure~\ref{fig:schematic}b).  

\begin{figure}[htbp]
\begin{minipage}[t]{80mm}
\hspace{-0.5cm}
    \epsfig{file=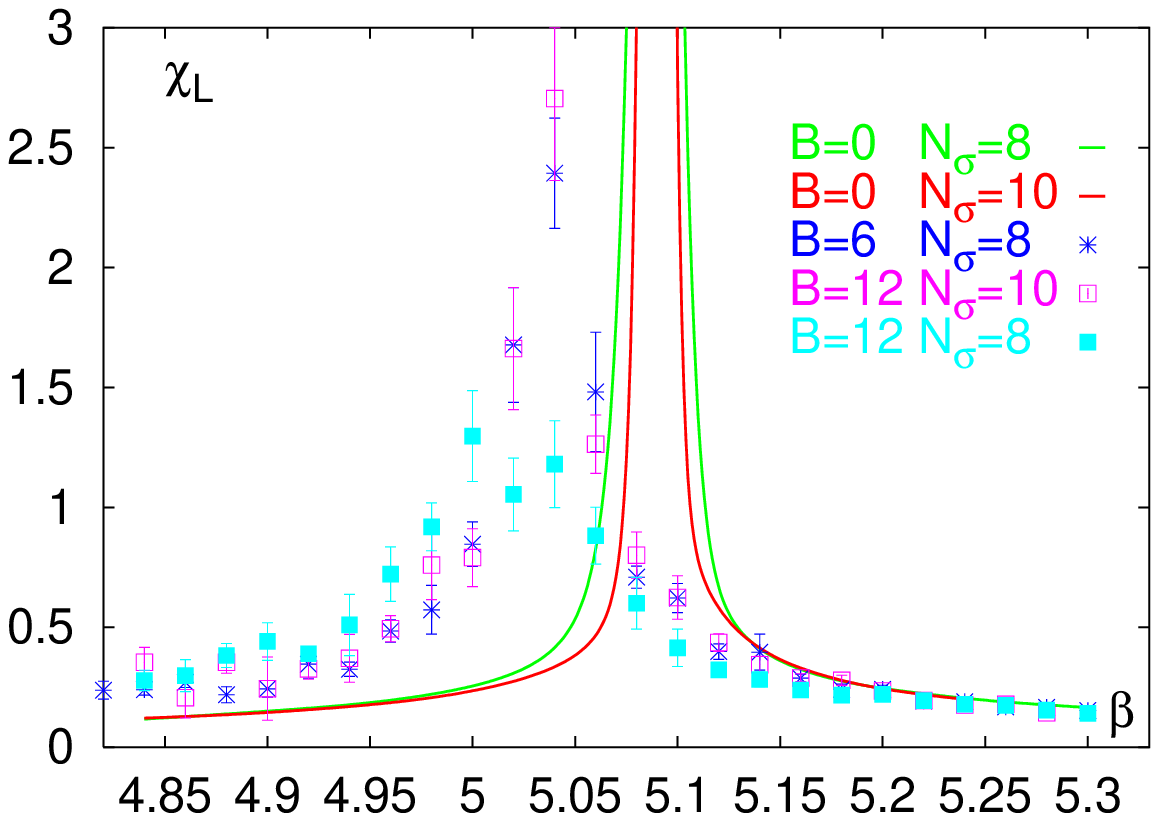, width=80mm}
  \vspace*{-5ex}
\end{minipage}
\begin{minipage}[t]{80mm}
\hspace{-0.9cm}
    \epsfig{file=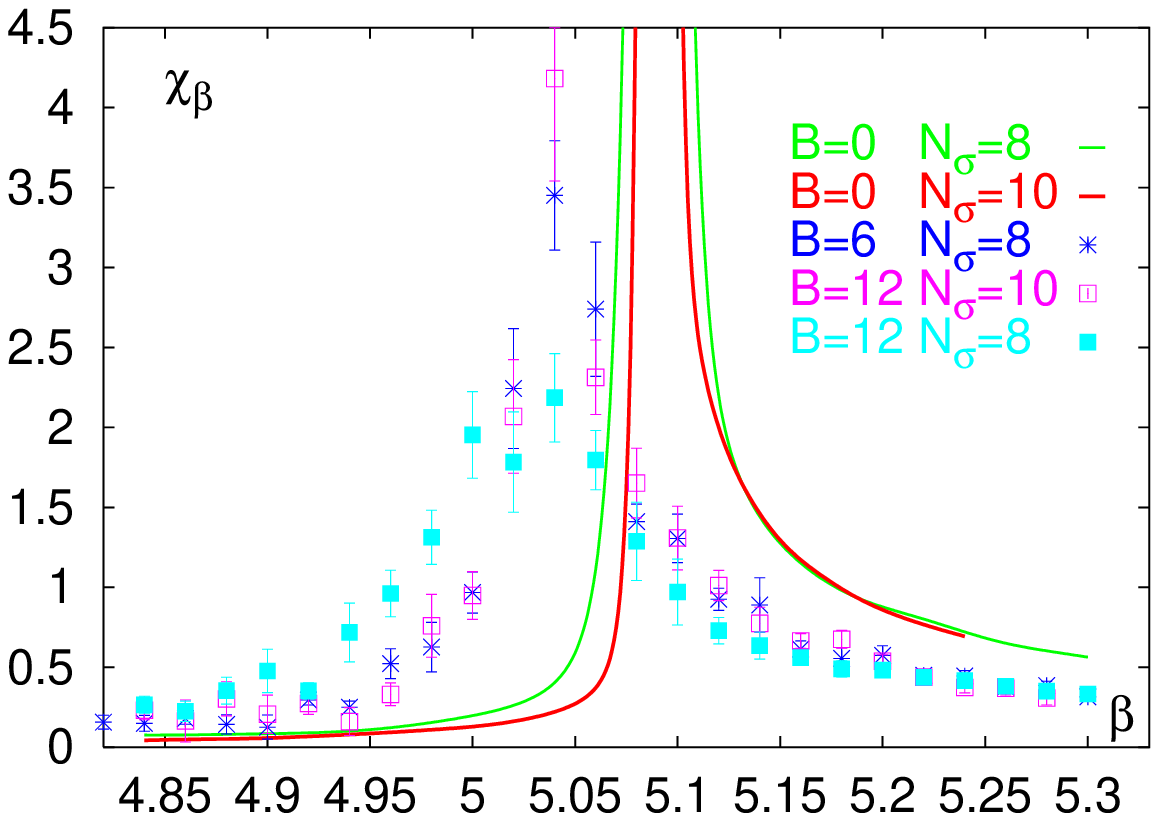, width=80mm}
\end{minipage}
  \caption{Polyakov loop susceptibility ($\chi_L$) and derivative of the Polyakov
loop with respect to $\beta$ ($\chi_\beta$) for different values of $B$
and lattices with spatial extent $N_\sigma =8$ and 10. For 
$B=0$ we only show the Ferrenberg-Swendsen interpolations of the data
for better visibility.}
  \label{fig:polsus}
\end{figure}

To gain further insight into the structure of this regime we also analyzed 
various susceptibilities.  In Figure~\ref{fig:polsus} we 
show the conventional Polyakov loop susceptibility,
\begin{equation}
\chi_L = N_\sigma^3 \biggl(\langle |L|^2 \rangle -\langle |L| \rangle^2\biggr)~~,
\label{chiL}
\end{equation}
as well as the derivative of $\langle |L| \rangle$ with respect to $\beta$,
\begin{equation}
\chi_\beta = {\partial \langle |L| \rangle \over \partial \beta} 
= \langle |L| \cdot S_G\rangle - \langle |L| \rangle \langle S_G\rangle~~.
\end{equation}
Both response functions
reflect the existence of a transition region that becomes broader with
increasing $n_B$. Compared to the behaviour at $B=0$ they also change 
continuously in this region. 
Such a behaviour might as well just correspond to
a smooth crossover to the high temperature regime;
a conclusion also drawn from the heavy quark simulations
with non-zero chemical potential \cite{Blum}. To establish the existence
of a region of coexisting phases with certainty will thus require a 
further, detailed analysis of finite size effects.

The width of the transition region gives some indication for the shift of the 
critical temperature when going from $B=0$ to a baryon number density 
which roughly corresponds to nuclear matter density. The gauge coupling
at which the Polyakov loop expectation value starts rising rapidly is shifted
from $\beta_c = 5.09$ at $n_B/T^3\equiv 0$ to $\beta_c \simeq 4.95$ 
at $n_B/T^3 = 0.032$.
A rough estimate based on the non-perturbative $\beta$-function for the 
$SU(3)$ gauge theory \cite{Boyd} suggests that this shift corresponds to 
a decrease of the critical temperature of about 15\%.

\begin{figure}[htbp]
  \begin{center}
    \leavevmode
    \epsfig{file=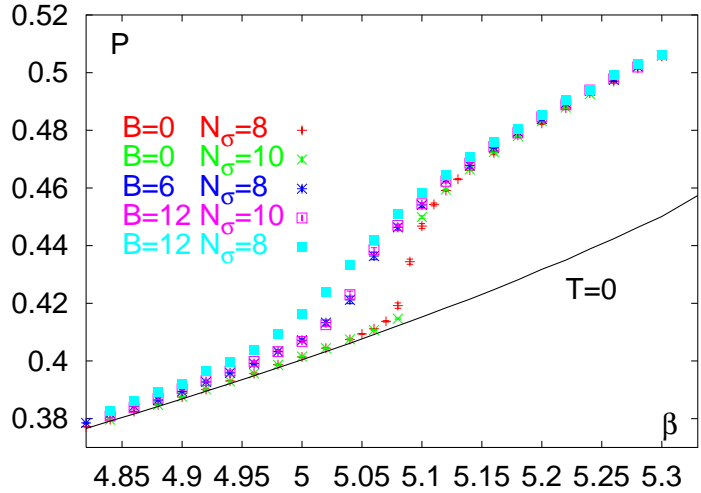, width=100mm}
  \end{center}
  \vspace*{-5ex}
  \caption{Plaquette expectation value for different values of $B$
and lattices with spatial extent $N_\sigma =8$ and 10.
The solid line shows a spline interpolation for the zero temperature 
plaquette expectation value, $P_0$, calculated on an $8^4$ lattice.}
  \label{fig:plaquette}
\end{figure}

The onset of deconfinement with increasing temperature is reflected in a 
sudden rise of bulk thermodynamic observables. For $B>0$ we thus expect to 
find a shift of this onset region to smaller temperatures. Otherwise,
however, we expect that observables like e.g. the free energy density
show a temperature dependence similar to that in the pure gauge sector. In the 
high temperature limit we expect to find a gluon gas slightly modified
due to the presence of static quarks.
    
\begin{figure}[htbp]
  \begin{center}
    \leavevmode
    \epsfig{file=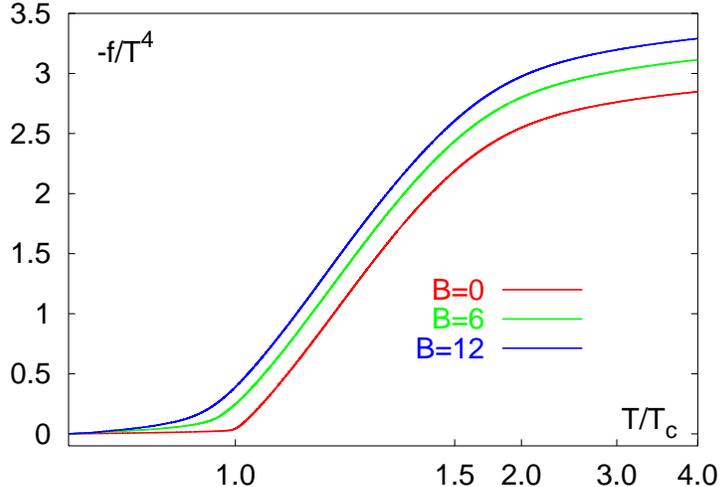, width=100mm}
  \end{center}
  \vspace*{-5ex}
  \caption{The negative of the free energy density in units of $T^4$ evaluated for 
different values of $B$ and lattices with spatial extent $N_\sigma =8$.
The abscissa is a linear scale in terms of the gauge
coupling $\beta$ covering the interval [4.74, 6.06]. We only show 
tick marks at the critical couplings on lattices with temporal 
extent $N_\tau=2,~3,~4,~6$ and 8, which correspond here 
to temperatures $T/T_c=1,~1.5,~2,~3$ and 4.}
  \label{fig:freeenergy}
\end{figure}

The free energy density can be calculated following the same approach used 
for $B=0$, \ie~through an integration over differences
of plaquette expectation values calculated on asymmetric ($8^3\times 2$)
and symmetric ($8^4$) lattices \cite{Boyd}. In Figure~\ref{fig:plaquette}
we show the plaquette expectation value,
\begin{equation}
P = {1\over 6 N_\sigma^3 N_\tau} \langle S_G \rangle~~,
\end{equation}
calculated for different values of $B$. The behaviour 
is similar to that of the Polyakov loop; with increasing $n_B$ the transition 
region broadens. The area between these data and the corresponding zero 
temperature results (solid line) increases. 
From this we obtain the 
free energy density\footnote{We define as zero temperature lattice a
symmetric lattice, $N_\tau = N_\sigma$. The plaquette expectation value
($P_0$) calculated on the zero temperature lattice is used for
normalization of the free energy.},
\begin{equation}
{f\over T^4}\Big\vert_{\beta_0}^{\beta} =~-
6 N_\tau^4 \int_{\beta_0}^{\beta}
{\rm d}\beta' \bigl[ P_0-P \bigr]~.
\label{freelat}
\end{equation}
As can be seen from Figure~\ref{fig:freeenergy} the free energy density
decreases at fixed temperature with increasing $B$. This results in a shift 
of the onset of
deconfinement to smaller temperatures. In the high temperature limit 
we indeed find that the free energy density is close to that of an
ideal gluon gas at $B=0$. The contribution of static quarks remains
small.

\section{Conclusions}
We have formulated the quenched limit of QCD at non-zero baryon number.
Although this formulation still leads to a path integral representation
of the partition function with an integrand that is not strictly positive
it can be handled quite well numerically for moderate values of the 
baryon number. A first numerical simulation of the gluon thermodynamics
in the background of static quarks shows the expected behaviour. We find
indications for a region of coexisting phases, which broadens with 
increasing baryon 
number density. The critical temperature, at which the system enters
this region, is shifted towards smaller temperatures with increasing
baryon number density. At high temperature we recover the physics of a
gluon gas similar to the $B=0$ case.

Although the transition clearly reflects the physics of deconfinement
it may not be a true thermal phase transition that leads to discontinuous
changes 
in thermodynamic observables. A clarification of this point will require 
calculations on larger lattices. If the statement turns out to be correct
one may have to think about the deconfinement transition at non-zero
baryon number in terms of a non-thermal percolation transition 
which anyhow seems to be a more appropriate physical picture for the 
deconfinement transition at zero temperature \cite{Satz,Celik}.

We also find (indirect) evidence that the heavy quark potential does tend
to a finite value at large distances already in the hadronic phase. The
increase of the Polyakov loop expectation value with increasing baryon number
density suggests that already at low temperatures string breaking starts
at short distances. The heavy quark potential thus may be significantly
modified in dense nuclear matter. 

The formulation of quenched QCD in the presence of static quarks seems to
be an appropriate model for the thermodynamics of QCD at non-zero baryon
number. It may be similarly useful for the analysis of thermal properties
of hadronic matter at non-zero density as the pure $SU(3)$ gauge theory
has been for the understanding of the finite temperature deconfinement 
transition. 

\paragraph{\bf Acknowledgments:}
This work was partly supported by the TMR network {\it Finite Temperature
Phase Transitions in Particle Physics}, EU contract no. ERBFMRX-CT97-0122
and the DFG through grant no. KA 1198/4-1.

\appendix
\section{Appendix}

In this appendix we will derive the representation of $\hat{f}_B$ given
in Eq.~\ref{partitions}
and give the explicit representation of $\hat{f}_B$ for a 
few values of $B$. 

The starting point for deriving Eq.~\ref{partitions} is the representation
of $\hat{f}_B$ given in Eq.~\ref{f_B_strong2},
\begin{eqnarray}
\hat{f}_B &=& (2 \kappa)^{B N_\tau}\sum_{X}
\prod_{k=1}^{B^{\prime}}
\bar{D}_{n_k} (\vec{x}_k) \quad ,
\label{append}
\end{eqnarray}
where $B^{\prime} \le B$ is the number of distinct sites $\vec{x}_k$ appearing
in $X$ and $n_{k}$ is the quark {\it occupation number}
for these distinct sites $\vec{x}_k$. They obey the
constraint $\sum_{k} n_{k} = B$ with $0 \le n_{k} \le 6n_f$. 

Let us set up some general rules for the evaluation of $\hat{f}_B$. Any term
in the sum appearing in Eq.~\ref{append} is characterized by $6n_f$ 
numbers $k=(k_1,..., k_{6n_f})$ where $k_i$ indicates how many sites are 
occupied $i$-times. The numbers $k_i$ are constrained by $\sum_i i k_i = B$. 
We thus may replace the vector $X=(\vec{x}_1,...,\vec{x}_B)$ by a vector 
$Y=(\vec{x}_1,...,\vec{x}_{B^{\prime}})$ which only lists the distinct sites
and is ordered according to the number of times these sites appear in $X$.
The additional vector $k$ keeps track of this information,
\ie~the first $k_1$ sites in $Y$ appear only once in $X$, from $k_1+1$
to $k_1+k_2$ we label the sites which appear twice in $X$ 
and so on. We then may write

\begin{eqnarray}
\hat{f}_B &=& (2 \kappa)^{B N_\tau}\sum_{Y,k} S_k(Y) \equiv 
(2 \kappa)^{B N_\tau} \sum_k \bar{S}_k \quad .
\label{f_B_strong_b}
\end{eqnarray}
with 
\begin{equation}
S_k (Y) \equiv S_{(k_1,...,k_{6n_f})} (Y) = \prod_{i=1}^{6n_f} 
\prod_{j=h_{i-1}+1}^{h_{i}} \bar{D}_{i}(\vec{x}_j)~~.
\label{sk}
\end{equation}
Here $h_0=0$ and $h_i = \sum_{1\le j \le i} k_j$ and thus $h_{6n_f}=B^{\prime}$.
Now we can perform the sum over $Y$ for a fixed 
set of occupation numbers for the $B^{\prime}$ distinct sites, which is 
given by the vector $k$,
\begin{equation}
\bar{S}_k  \equiv \sum_Y~^{\hspace{-0.2truecm} \prime}~ S_k (Y)~~.
\end{equation}
The prime on the sum reminds us that in doing 
this sum we have to avoid, of course, double counting. A pair
of sites $\vec{x}_i \ne \vec{x}_j$ which have identical occupation 
numbers  should appear only once in the sum, \ie~interchanging $\vec{x}_i$ 
and $\vec{x}_j$ should be counted as one configuration. This means that all
$Y$ which only differ by a permutation within an interval 
$(h_{i}+1,h_{i+1})$ are equivalent. We can give up this restriction and 
divide by appropriate factors,
\begin{equation}
\bar{S}_k  \equiv \biggl( \prod_{l=1}^{6n_f} k_l! \biggr)^{-1} \sum_Y S_k (Y)~~.
\label{sk_tot}
\end{equation}
Now we only have to insure that all $\vec{x}_i \ne \vec{x}_j$. We also want
to eliminate this restriction and do independent sums over all $\vec{x}_i$.
This is achieved by performing one of the sums over $\vec{x}_i$ without any
restriction and at the same time correct for the constraint by subtracting
a term where two summation indices have been contracted. This process of 
factorization and contraction is repeated $(B^{\prime}-1)$ times.
Let us illustrate this by doing the first step explicitly:

\begin{eqnarray}
\sum_Y S_k (Y) &=& \sum_{\vec{x}_1\ne \vec{x}_2...\ne \vec{x}_{B^{\prime}} }
\prod_{j=1}^{B^{\prime}} \bar{D}_{n_j}(\vec{x}_j) \nonumber \\
&=& \biggl( \sum_{\vec{x}_1} \bar{D}_{n_1}(\vec{x}_1) \biggr) \biggl(
\sum_{\vec{x}_2...\ne \vec{x}_{B^{\prime}} }
\prod_{j=2}^{B^{\prime}} \bar{D}_{n_j}(\vec{x}_j) \biggr) - \nonumber\\
& &
\sum_{t=2}^{B^{\prime}} 
\sum_{\vec{x}_2...\ne \vec{x}_{B^{\prime}} } 
\bar{D}_{n_1}(\vec{x}_t)\prod_{j=2}^{B^{\prime}} 
\bar{D}_{n_j}(\vec{x}_j) \nonumber\\
&=& [ \bar{D}_{n_1} ] 
\sum_{\vec{x}_2...\ne \vec{x}_{B^{\prime}} }
\prod_{j=2}^{B^{\prime}} \bar{D}_{n_j}(\vec{x}_j) -
\sum_{t=2}^{B^{\prime}}
\sum_{\vec{x}_2...\ne \vec{x}_{B^{\prime}} }
\bar{D}_{n_1}(\vec{x}_t)\prod_{j=2}^{B^{\prime}}
\bar{D}_{n_j}(\vec{x}_j)  \quad , \nonumber      
\end{eqnarray}
where $[... ]$ denotes the sum over the lattice taken 
on a single configuration, \ie~
\begin{equation}
\biggl[\prod_i\bar{D}_{n_i}\biggr] = 
\sum_{\vec{x}} \prod_i\bar{D}_{n_i} (\vec{x})~~.
\label{brackets}
\end{equation} 
When continuing this process of factorization and contraction we 
arrive at a cluster decomposition of the $B^{\prime}$ factors we 
started with. Contracting two summation indices leads to a factor $(-1)$.
We thus obtain a representation of $\hat{f}_B$ as a sum over 
products of clusters 
\begin{equation}
F(\alpha) \equiv F(a_1,...,~a_{6n_f}) = (-1)^{(j - 1)} 
\bigl( j - 1\bigr)! 
\biggl[\prod_{l=1}^{6n_f} {\bar{D}_{l}^{a_l} \over a_l !} \biggr]\quad ,
\end{equation}
with 
\begin{equation}
j=\sum_{l=1}^{6n_f} a_l \quad .
\end{equation}
Here 
we have explicitly given the combinatorial factor for a cluster
of length $j$, \ie~ a factor $(-1)$ for each of the $(j-1)$
contractions needed to generate a cluster of length $j$ and a factor $(j-1)!$
for the number of ways one can create such a cluster. We also included
a combinatorial factor that takes into account that the permutation of $a_l$
factors $D_l$ appearing in a cluster does not lead to a new cluster
decomposition. 

In total there are $k_l!$ possibilities to distribute the
various terms $D_l$. However, we have to take into account that each cluster 
can appear several times, \ie~ its degeneracy 
is $g_{(a_1,...,~a_{6n_f})}$. Also the permutation of identical clusters
does not lead to a new cluster decomposition.   
We thus arrive at the representation
\begin{equation}
\sum_Y S_k (Y) = 
\biggl( \prod_{l=1}^{6n_f} k_l! \biggr)
\sum_{\{g_\alpha\}}\delta(\{g_\alpha\})
\prod_{\{\alpha \}} \biggl({1\over g_\alpha!} F(\alpha)^{g_\alpha}\biggr)
\quad .
\label{gsum}
\end{equation}
where $\alpha \equiv (a_1,...~a_{6n_f})$ and the product runs over the set of 
vectors ${\alpha}$ which satisfy
\begin{equation}
\sum_{l=1}^{6n_f} l a_l \le B 
\quad .
\label{constraintc}
\end{equation}
Associated with
each contributing vector $\alpha$ is a sum over the non-negative integers
$g_\alpha$ which is symbolically represented in (\ref{gsum}) by the sum 
over $\{g_\alpha \}$, 
\begin{equation}
 \sum_{\{g_\alpha \}} \equiv
\sum_{g_{(1,0,...,0)}=0}^B \sum_{g_{(2,...,0)}=0}^{{\rm int}(B/2)} ...
\sum_{g_\alpha =0}^{i(\alpha)} ...  \quad ,
\label{gsumb}
\end{equation}
where $i(\alpha)$ denotes the integer part of $B/\sum_{l=1}^{6n_f} l a_l$.
The Kronecker-$\delta$, $\delta(\{ g_\alpha \})$, appearing in (\ref{gsum})
summarizes the constraints
the set of summation indices has to satisfy,
\begin{equation}
\delta(\{ g_\alpha \}) = \delta\biggl( B- \sum_{l=1}^{6n_f} l k_l \biggr)
\prod_{l=1}^{6n_f} \delta\biggl( k_l-\sum_{\{\alpha \}} g_\alpha a_l \biggr)
\quad .
\label{constraintg}
\end{equation}
We note that the combinatorial factor appearing in (\ref{gsum}) in front of 
the sums will just cancel the factor appearing in (\ref{sk_tot});
in fact, they do no longer depend on the actual choice of the
vector $k$. We thus may easily sum over all possible vectors $k$ 
which yields essentially the result given in (\ref{gsum}) with a less
restrictive constraint on the possible choice of the summation indices 
$\{g_\alpha \}$, 
\begin{equation}
\hat{f}_B = (2 \kappa)^{B N_\tau}
\sum_{\{g_\alpha \}} \delta\biggl( B- \sum_{l=1}^{6n_f} 
\sum_{\{\alpha \}} l g_\alpha a_l 
\biggr)
\prod_{\{\alpha \}} \biggl({1\over g_\alpha !} F(\alpha)^{g_\alpha}\biggr)
\quad ,
\end{equation}
where the allowed set of numbers $\{ \alpha \}$ is constrained only by
(\ref{constraintc}).  

This is the representation of $\hat{f}_B$ given in Eq.~\ref{partitions}.
In the following we will give the explicit representation for a few small
values of $B$. 

\subsection{B=3}

We get 3 contributions to $\hat{f}_3$:

\begin{equation}
\hat{f}_3 = (2 \kappa)^{3 N_\tau} \biggl(\bar{S}_{001000} + \bar{S}_{110000} + 
\bar{S}_{300000} \biggr) \quad .
\end{equation}
The terms $\bar{S}_k$ are sums over products of terms $D_i$ which are
defined in Eq.~\ref{Di}. The three different 
contributions are given in Table~\ref{tab:states3}.
\begin{table}
\begin{center}
\begin{tabular}{|c|l|c|}\hline
$k$&$S_k(Y)$&$\prod k_i !$ \\
\hline
\hline
001000&$\bar{D}_3(x_1)$& 1\\
110000&$\bar{D}_1(x_1) \bar{D}_2(x_2)$&1 \\
300000&$\bar{D}_1(x_1) \bar{D}_1(x_2) \bar{D}_1(x_3)$& 3!\\
\hline
\end{tabular}
\end{center}
\caption{3-quark contributions}
\label{tab:states3}
\end{table}
We thus obtain 
\begin{equation}
\hat{f}_3 = (2 \kappa)^{3 N_\tau} \biggl( 
[\bar{D}_3] -[\bar{D}_1 \bar{D}_2] + 
{1\over 3} [\bar{D}_1^3] + 
 [ \bar{D}_1] ([\bar{D}_2] - {1\over 2} [\bar{D}_1^2])   +
{1\over 6} [\bar{D}_1]^3 \biggr)~~  .
\end{equation}

\subsection{B=6}

For $B=6$ one gets 11 contributions to $\hat{f}_6$:
\begin{equation}
\hat{f}_6 = (2 \kappa)^{6 N_\tau} \sum_k \bar{S}_k \quad .
\end{equation}
The different 6-vectors $k$ and the corresponding contributions $S_k(Y)$ are 
listed in the Table~\ref{tab:states6}. 

\begin{table}
\begin{center}
\begin{tabular}{|c|l|c|}\hline
$k$&$S_k(Y)$&$\prod k_i !$ \\
\hline
\hline
000001&$\bar{D}_6(x_1)$& 1\\
100010&$\bar{D}_1(x_1) \bar{D}_5(x_2)$&1 \\
010100&$\bar{D}_2(x_1) \bar{D}_4(x_2)$& 1\\
002000&$\bar{D}_3(x_1) \bar{D}_3(x_2)$ & 2!\\
200010&$\bar{D}_1(x_1) \bar{D}_1(x_2) \bar{D}_4(x_3)$& 2! \\
111000&$\bar{D}_1(x_1) \bar{D}_2(x_2) \bar{D}_3(x_3)$& 1\\
030000&$\bar{D}_2(x_1) \bar{D}_2(x_2) \bar{D}_2(x_3)$& 3!\\
301000&$\bar{D}_1(x_1) \bar{D}_1(x_2) \bar{D}_1(x_3) \bar{D}_3(x_4)$& 3! \\
220000&$\bar{D}_1(x_1) \bar{D}_1(x_2) \bar{D}_2(x_3) \bar{D}_2(x_4)$& 2! 2!\\
410000&$\bar{D}_1(x_1) \bar{D}_1(x_2) \bar{D}_1(x_3) \bar{D}_1(x_4) 
\bar{D}_2(x_5)$& 4! \\
600000&$\bar{D}_1(x_1) \bar{D}_1(x_2) \bar{D}_1(x_3) \bar{D}_1(x_4) 
\bar{D}_1(x_5) \bar{D}_1(x_6)$& 6!\\
\hline
\end{tabular}
\end{center}
\caption{6-quark contributions}
\label{tab:states6}
\end{table}
This again gives rise to contributions, which can be ordered according 
to the number of clusters contributing   
\begin{equation}
\hat{f}_6 = (2 \kappa)^{6 N_\tau} \sum_{i=1}^{6} a_6(i) ~~,
\end{equation}
with 
\begin{eqnarray}
a_6(1) &=&
-{1\over 6} [\bar{D}_1^6] + [\bar{D}_1^4 \bar{D}_2] - {3\over 2} [\bar{D}_1^2 \bar{D}_2^2] + {1\over 3}[\bar{D}_2^3] - 
   [\bar{D}_1^3 \bar{D}_3] + \nonumber \\
& &2 [\bar{D}_1 \bar{D}_2 \bar{D}_3] - {1\over 2} [\bar{D}_3^2] + [\bar{D}_1^2 \bar{D}_4] - [\bar{D}_2 \bar{D}_4] - 
   [\bar{D}_1 \bar{D}_5] + [\bar{D}_6] \\
a_6(2) &=&
{1\over 18} [\bar{D}_1^3] ([\bar{D}_1^3]  - 6 [\bar{D}_1 \bar{D}_2]) + 
     {1\over 2} [\bar{D}_1 \bar{D}_2]^2 + \nonumber \\
& & [\bar{D}_3] ({1\over 3} [\bar{D}_1^3] - [\bar{D}_1 \bar{D}_2] + {1\over 2} [\bar{D}_3] )  - \nonumber \\ 
& &{1\over 2} (2 [\bar{D}_2] - [\bar{D}_1^2]) ({1\over 2} [\bar{D}_2^2] + {1\over 4} [\bar{D}_1^4] - 
      [\bar{D}_1^2 \bar{D}_2] + [\bar{D}_1 \bar{D}_3] -[\bar{D}_4] )+ \nonumber \\
& &   [\bar{D}_1] ({1\over 5} [\bar{D}_1^5] - [\bar{D}_1^3 \bar{D}_2] + [\bar{D}_1 \bar{D}_2^2] + 
      [\bar{D}_1^2 \bar{D}_3] - [\bar{D}_2 \bar{D}_3] - \nonumber \\
& &   [\bar{D}_1 \bar{D}_4] + [\bar{D}_5]) \\
a_6(3) &=& {1\over 48} (2 [\bar{D}_2] - [\bar{D}_1^2] )^3   + 
\nonumber \\
& &   {1\over 6} [\bar{D}_1] (2 [\bar{D}_2] - [\bar{D}_1^2] ) ([\bar{D}_1^3] - 3 [\bar{D}_1 \bar{D}_2] + 3 [\bar{D}_3])-
\nonumber \\
& & {1\over 2} [\bar{D}_1]^2 ({1\over 2} [\bar{D}_2^2] + {1\over 4} [\bar{D}_1^4] - [\bar{D}_1^2 \bar{D}_2] +  
      [\bar{D}_1 \bar{D}_3] - [\bar{D}_4]) \\
a_6(4) &=& {1\over 16} [\bar{D}_1]^2 (2 [\bar{D}_2] - [\bar{D}_1^2] )^2 + 
   {1\over 6} [\bar{D}_1]^3 ({1\over 3} [\bar{D}_1^3] - [\bar{D}_1 \bar{D}_2] + [\bar{D}_3]) \\
a_6(5) &=& {1\over 48} [\bar{D}_1]^4 (2 [\bar{D}_2] - [\bar{D}_1^2] ) \\
a_6(6) &=& {1\over 720} [\bar{D}_1]^6 \quad .
\end{eqnarray}

These coefficients have been generated with Mathematica. In total there
are 58 different terms contributing to $\hat{f}_{6}$. It is apparent that 
the number of terms increases rapidly with $B$. For $B=12$ and $n_f=1$ there 
are 58 different vectors $k$ which give rise to 2739 terms in $\hat{f}_{12}$.  

\baselineskip 15pt

%

\begin{thebibliography}{99}
\baselineskip 10pt
\bibitem{Joswig}
J. Engels, R. Joswig, F. Karsch, E. Laermann, M. L\"utgemeier
and B. Petersson, Phys. Lett. B396 (1997) 210. 
\bibitem{Has83}P. Hasenfratz and F. Karsch, Phys. Lett. 125B (1983) 308.
\bibitem{Kog83}J. Kogut, M. Matsuoka, M. Stone, H.W. Wyld, J.H. Shenker,
J. Shigemitsu and D.K. Sinclair, Nucl. Phys. B225 [FS9] (1983) 93. 
\bibitem{Barxx}I. Barbour, N.-E. Behilil, E. Dagotto, F. Karsch, A. Moreo, 
M. Stone and H.W. Wyld, Nucl. Phys. B275 [FS17] (1986) 296. 
\bibitem{Stexx} M.A. Stephanov, Phys. Rev. Lett. 76 (1996) 4472.    
\bibitem{Bender}I. Bender, T. Hashimoto, F. Karsch, V. Linke, A. Nakamura,
M. Plewnia, I.O. Stamatescu and W. Wetzel, Nucl. Phys. B (Proc. Suppl.) 26 
(1992) 323.
\bibitem{Blum} T.C. Blum, J.E. Hetrick  and D. Toussaint, Phys. Rev. Lett.
76 (1996) 1019.
\bibitem{Ker89}J. Kert\'esz, Physica A161 (1989) 58.
\bibitem{Satz}H. Satz, Nucl. Phys. A642 (1998) 130c.
\bibitem{Red87}D.E. Miller and K. Redlich, Phys. Rev. D35 (1987) 2524.
\bibitem{Weixx}A. Roberge and N. Weiss, Nucl. Phys. B275 [FS17] (1986) 734. 
\bibitem{Bar98}I.M. Barbour, S.E. Morrison, E.G. Klepfish, J.B. Kogut
and M.-P. Lombardo, Nucl. Phys. B (Proc. Suppl.) 60 (1998) 220. 
\bibitem{MDP} F. Karsch and K.-H. M\"utter, Nucl. Phys. B313 (1989) 541.
\bibitem{Alf98}M. Alford, A. Kapustin and F. Wilczek , 
Phys. Rev. D59 (1999) 054502;\\
M. Alford, New Possibilities for QCD at finite Density, 
IASSNS-HEP-98-81, hep-lat/9809166. 
\bibitem{Matt}  T. Matthews and A. Salam, Nuovo Cim. 12 (1954) 563; 2 (1955) 120;\\
C. Lang and H. Nicolai, Nucl. Phys. B200 (1982) 135.
\bibitem{Engels} B. Berg, J. Engels, E. Kehl, B. Waltl and H. Satz,
Z. Phys. C31 (1986) 167. 
\bibitem{Wilson} K.G. Wilson, Phys. Rev. D10 (1974) 2445.
\bibitem{DeTar}C. DeTar, O. Kaczmarek, F. Karsch and E. Laermann,
Phys. Rev. D59 (1999) 031501.
\bibitem{Boyd}G. Boyd, J. Engels, F. Karsch, E. Laermann, C. Legeland, 
M. L\"utgemeier and B. Petersson, Nucl. Phys. B469 (1996) 419.
\bibitem{Celik}T. Celik, F. Karsch and H. Satz, Phys. Lett. 97B (1980) 128.
\end{thebibliography}
\end{document}